\documentclass[aps,prd,superscriptaddress,12pt,showpacs,notitlepage]{revtex4}
\usepackage{amsfonts}
\usepackage{graphics}
\usepackage{graphicx}

\newcommand{\be}{\begin{equation}}
\newcommand{\ee}{\end{equation}}
\newcommand{\bea}{\begin{eqnarray}}
\newcommand{\eea}{\end{eqnarray}}
\newcommand{\pa}{\partial}
\newcommand{\bb}{\bibitem}
\begin{document}
\title{Supersymmetric extensions of k-field models}
\author{D. Bazeia}
\affiliation{Departamento de F\'{\i}sica, Universidade Federal da Para\'{\i}ba, 58051-970 Jo\~ao Pessoa, Para\'{\i}ba, Brazil}
\author{R. Menezes}
\affiliation{Departamento de Ci\^{e}ncias Exatas, Universidade Federal da Para\'\i ba,\\ 58297-000 Rio Tinto, Para\'{\i}ba, Brazil}
\author{A. Yu. Petrov}
\affiliation{Departamento de F\'{\i}sica, Universidade Federal da Para\'{\i}ba, 58051-970 Jo\~ao Pessoa, Para\'{\i}ba, Brazil}


\begin{abstract}
We investigate the supersymmetric extension of k-field models, in which the scalar field is described by generalized dynamics. We illustrate some results with models that support static solutions with the standard kink or the compact profile.  
\end{abstract}

\maketitle 

Topological defects may have important consequences in a diversity of contexts, in particular in Cosmology, where they appear very naturally through the presence of phase transitions in the early universe \cite{V}. As it is well-known, one of the simplest structures, domain walls, spring from the breaking of discrete symmetry, guided by scalar fields, which will be the main focus of the present work. In this case, in models describing spontaneous breaking of some discrete symmetry, the presence of topological defects in canonical models of scalar fields with standard dynamics, the equations of motion are shown to reduce to first-order differential equations, with the model being nothing but the bosonic portion of some more general supersymmetric theory. One usually refers to this possibility as the first-order framework, where the topological defects appear as solutions of first-order equations as BPS states, named after the pioneer investigations of Bogomol'nyi, and Prasad and Sommerfield \cite{BPS}. 

With the recent discovery that the universe is evolving through an accelerated expansion, models described by scalar fields with non-standard dynamics are being studied in the cosmological context, for instance, as the so-called k-fields, first introduced in the context of inflation \cite{m1} and then as k-essence models, suggested to solve the cosmic coincidence problem \cite{m2}. { We recall that k-inflation \cite{m1} has appeared as a way to reconcile the string dilaton with inflation. However, as one knows, supersymmetry is supposed to be unbroken
at high energies, so it may affect the inflationary evolution. In this sense, supersymmetric extensions of k-field dynamics seem to be of current interest to the physics at high energy scales.} 

{The case of scalar fields with standard dynamics has been studied in several different contexts, and here we point out a diversity of investigations on the presence of domain walls in supergravity theories -- see, e.g., Ref.~{\cite{mc}} and references therein. On the other hand,} the presence of scalar field models with non-standard dynamics has stimulated several recent investigations, dealing with properties of topological structures in this new scenario \cite{B,b1,s1,o,s2,b2,s3}. The main issue here is to change the scalar field dynamics from the standard one to some generalized dynamics, also known as k-fields or $f(X)$ models, with $X=\pa^\mu\phi\pa_\mu\phi$, to accommodate some distinct possibilities, including the presence of compactons, e.g., of classical static solutions with compact support \cite{c}. 

In particular, in the works \cite{b2} some of us have shown how to describe the presence of defect structures in models described by scalar fields with generalized dynamics within a first-order framework. Thus, since we know that under standard dynamics, the presence of first-order equations in general indicates the bosonic portion of some supersymmetric theory, in the present work we then study supersymmetric extensions of scalar field models with generalized dynamics. The subject is of current interest, since the presence of supersymmetry may ease calculation and may lead us to scenarios compatible with superstring theory. {We believe that the idea we explore below may contribute as alternative for the standard inflationary evolution of the early Universe \cite{m1,m2} and also, to the domain wall/braneworld scenario \cite{mc,b1,b2,CGB}.}

The search for exact solutions is a very important direction of studies in theoretical physics, and has led to different methods for finding analytical solutions. Among different methods of finding the solutions, special role is played by the first-order framework where the exact solutions are found through the reduction of the equations of motion to appropriate first-order differential equations. This method has been efficiently applied to different field theory models, with direct interest into recent cosmological, gravitational and braneworld issues \cite{CGB}. These investigations have motivated us to extend the first-order framework to the context of supersymmetric field theory models. The supersymmetry is now treated as a fundamental symmetry of the nature \cite{SGRS}, the perturbative approach for the supersymmetric models is well-developed, both in three-\cite{my3} and four-dimensional space-time \cite{ours}, therefore the interesting problem is the nonperturbative study of these model, in particular, answering the question whether the known exact solutions of the scalar field theory models admit supersymmetric extensions. 

{In this work we investigate the supersymmetric extension of scalar field models with generalized dynamics, which we refers to as the k-field models. As we are going to show, models recently investigated in \cite{b1,b2} support supersymmetric extensions, which are explicitly constructed below. In particular, we also show that some of the supersymmetric extensions which we construct admit solutions with compact support. In this sense, here we are giving the first step to learn how to bring braneworld with compact support to the supersymmetric environment. Evidently, the existence of structures of compact and non-compact support in k-field models in the presence of supersymmetry is of current interest, since they may contribute to modify the nature of the spacetime in such backgrounds \cite{m1,m2,mc,B,b1,s1,o,s2,s3,b2,CGB}.}

To start with, let us introduce the three-dimensional scalar superfield $\Phi(z)=\Phi(x,\theta)$ (the same structure takes place for two-dimensional scalar superfield as well, since spinor representations of the Lorentz group in two- and three-dimensional space-times are similar) \cite{SGRS}:
\bea
\Phi(x,\theta)=\phi(x)+\theta^{\alpha}\psi_{\alpha}(x)-\theta^2 F(x).
\eea
Its components can be defined as the following projections:
\bea
\label{comp}
\phi(x)=\Phi(z)|,\quad\,\psi_{\alpha}(x)=D_{\alpha}\Phi(z)|,\quad\,F(x)=D^2\Phi(z)|,
\eea
where the sign $|$ denotes that after differentiation the $\theta^{\alpha}$ in the expansion of the corresponding superfield are set to zero. Here we use notations of \cite{SGRS}, with the signature $diag(-++)$.

The simplest example of the superfield action in three-dimensional superspace, that is, the action for the models involving only scalar superfields, has the usual form (cf. \cite{my3}):
\bea
\label{scaf}
S=\int d^5 z\, \left[-\frac{1}{4}D^{\alpha}\Phi D_{\alpha}\Phi+f(\Phi)\right],
\eea
where $f(\Phi)$ is an arbitrary function of the scalar superfield but not of its derivatives. The case of the complex scalar superfield does not essentially differ.

The component form of this action is defined from the expression
\bea
\label{scaf1}
S=\int d^3x D^2 \left[-\frac{1}{4}D^{\alpha}\Phi D_{\alpha}\Phi+f(\Phi)\right]\Big|.
\eea
Here we have taken into account that $D^2=\frac{1}{2}D^{\alpha}D_{\alpha}$ \cite{SGRS}.
After straightforward differentiation, this expression is reduced to
\bea
S&=&\int d^3 x
\Big[\frac{1}{2}F^2-\frac{1}{2}i\psi_{\alpha}\pa^{\alpha\beta}\psi_{\beta}-
\frac{1}{2}\pa_\mu\phi\pa^\mu \phi+ 
\frac{1}{2}f_{\phi\phi}(\phi)\psi^{\alpha}\psi_{\alpha}+f_{\phi}(\phi)F\Big].
\eea
The auxiliary field $F$ can be eliminated with use of its equation of motion $F=-f_{\phi}(\phi)$; this gives
\bea
\label{simp}
S&=&\int d^3 x
\Big[-\frac{1}{2}(f_{\phi}(\phi))^2-\frac{1}{2}i\psi_{\alpha}\pa^{\alpha\beta}\psi_{\beta}-
\frac{1}{2}\pa_\mu\phi\pa^\mu \phi+\frac{1}{2}f_{\phi\phi}(\phi)\psi^{\alpha}\psi_{\alpha}\Big].
\eea
This is the standard case and for $f(\Phi)=\lambda\Phi^4$ (this is the highest possible renormalizable self-coupling of the scalar superfield; remind that the mass dimension of $\Phi$ is $1/2$, as well as of the derivative $D_{\alpha}$, and the dimension of $d^5 z$ is $-2$) we get the coupling $\lambda\phi^6$.

As a first nonstandard and instructive example let us consider the three-dimensional superfield theory with the following superfield action
\bea
\label{sact}
S_1=-\frac{1}{4}\int d^5zD^{\alpha}\Phi D_{\alpha}\Phi f(\pa^\mu\Phi\pa_\mu\Phi), 
\eea
where $\Phi$ is a three-dimensional real scalar superfield defined above, whose components are defined via (\ref{comp}). 

To obtain the bosonic sector of the action (\ref{sact}), we integrate over $\theta$ using equivalence between integration over Grassmannian variables and differentiation with respect to them
\bea
S_1=-\frac{1}{4}\int d^3x D^2[D^{\alpha}\Phi D_{\alpha}\Phi f(\pa^\mu\Phi\pa_\mu\Phi)]|.
\eea
It is clear that the only term depending on the bosonic components of the superfield $\Phi$ is
\bea
S_1=\frac{1}{4}\int d^3x (D^{\beta}D^{\alpha}\Phi) (D_{\beta}D_{\alpha}\Phi) f(\pa^\mu\Phi\pa_\mu\Phi)|.
\eea
Indeed, all other terms involve at least one term with only one derivative of $\Phi$, that is, they are fermion dependent.
Then, we use the well-known relation $D_{\beta}D_{\alpha}=i\pa_{\beta\alpha}-C_{\beta\alpha}D^2$, which yields
\bea
\label{acf1}
S_1=-\frac{1}{2}\int d^3x (\pa^\mu\phi\pa_\mu\phi-FF) f(\pa^\nu\phi\pa_\nu\phi),
\eea
where we used the identities $\pa_{\beta\alpha}=\pa_\mu\gamma^\mu_{\beta\alpha}$ and ${\rm tr}(\gamma^\mu\gamma^\nu)=2\eta^{\mu\nu}$.

We can eliminate the auxiliary field $F$ via its equation of motion
\bea
F\cdot f(\pa^\mu\phi\pa_\mu\phi)=0,
\eea 
whose solution is $F=0$; otherwise, the action would be zero. Thus, we arrive at the action
\bea
S=-\frac{1}{2}\int d^3x (\pa^\mu\phi\pa_\mu\phi) f(\pa^\nu\phi\pa_\nu\phi)=-\frac{1}{2}\int d^3x Xf(X),
\eea
where $X=\pa^\mu\phi\pa_\mu\phi$. We then succeeded to prove that the action (\ref{sact}) represents a supersymmetric extension of the k-field action
\bea
S_1=-\frac{1}{2}\int d^3x \tilde{f}(X),
\eea
with $\tilde{f}(X)=Xf(X)$. However, it only contains derivative of the scalar field, and so we need to go further to get more general models.

To do this, let us note that adding to the action (\ref{sact}) the potential term
\bea
\label{pot1}
\int d^5 z h(\Phi)=\int d^3x h_{\phi}(\phi)F,
\eea
and subsequent elimination of the auxiliary $F$ field yields
\bea\label{AAAA}
S_1^b=\frac{1}{2}\int d^3x \left(-Xf(X)-\frac{(h_{\phi})^2(\phi)}{f(X)}\right).
\eea
We note that the second term is not potential-like in general; indeed, it is not a function of $\phi$ alone. Thus the action of the form
\bea
S_0=-\frac{1}{2}\int d^3x (\tilde{f}(X)+h(\phi)),
\eea
admits the supersymmetric extension only either for $h(\phi)=0$ or for $\tilde{f}(X)=X$. The second case leads to the standard situation, in which $f(X)=1$, and this exactly reproduces the bosonic sector of (\ref{simp}), as expected.

Returning to the action (\ref{AAAA}), it is interesting to see that the corresponding equation of motion has the form 
\be
\partial_\mu \left(\left[f(X)+Xf^\prime - \frac{f^\prime}{f^2} h_\phi^2\right]\partial^\mu\phi\right)=\frac{h_\phi h_{\phi\phi}}{f(X)}.
\ee
Within this paper we will mostly deal with solutions dependent on only one spatial coordinate, say $x\equiv x_1$. So, for static solutions we can write 
\be
\left(\left[f(X)+Xf^\prime - \frac{f^\prime}{f^2} h_\phi^2\right]\phi^\prime\right)^\prime=\frac{h_\phi h_{\phi\phi}}{f(X)},
\ee
where $X=\phi^{\prime2}$ for static solutions. The energy density $\rho(x)=T^0_0$ is
\be\label{rhoEq}
\rho(x)=\frac12 Xf(X)+\frac12\frac{h_\phi^2}{f(X)}.
\ee 
The equation of motion can be integrated to give 
\be\label{eqeq}
-f^2(X) X + h_\phi^2 = \frac{2\,C f^2(X)}{f(X)+2Xf^\prime(X)}.
\ee 
Topological solutions usually appear for $C=0$, as one can see from Ref.~\cite{b1}. In this case we have
\be
f^2(X) X = h_\phi^2,
\ee
or
\be\label{firstorder}
\phi^\prime f(\phi^{\prime2})=\pm {h_\phi},
\ee
which naturally leads to the Bogomol'nyi bound. To see this, we use the eq.~(\ref{rhoEq}) to write the energy as
\be
E=\frac12 \int^\infty_{-\infty}dx \,\left(\phi^\prime \sqrt{f(\phi^{\prime2})} \mp \frac{h_\phi}{\sqrt{f(\phi^{\prime2})}}\right)^2
\pm \int^\infty_{-\infty}dx \frac{dh}{dx}.
\ee
Thus, the energy is minimized to the value $E=|h(\phi(\infty))-h(\phi(-\infty))|$ for solutions which solve the first-order equations (\ref{firstorder}). Another interesting feature of the solutions of the first-order equations (\ref{firstorder}) is that they make the two terms in the energy density in (\ref{rhoEq}) to contribute evenly, allowing the energy density to be written as
\be
\rho(x)=\phi^{\prime2}f(\phi^{\prime2})=h^2_\phi/f(\phi^{\prime2}).
\ee 
As one knows, the above are general features of the BPS states \cite{BPS}, that is, of the solutions that solve first-order differential equations.
 
To illustrate the general situation, let us choose $f(X)$ as the function $f(X)=|X|^{n-1}$. In this case, the equation (\ref{eqeq}) becomes
\be
\phi^{\prime(2n-2)} \left(\phi^{\prime(2n)} + \frac{2C}{2n-1}\right) =h_\phi^2,
\ee 
which is an algebraic equation in $\phi^\prime$. For $C=0$, it reduces to
\be
\phi^\prime = \pm h_\phi^{\frac1{2n-1}},
\ee
and the energy density (\ref{rhoEq}) has now the form $\rho(x)=h_\phi^{\frac{2n}{2n-1}}$. Here we take $n=2$ and as a first example, we consider the function
\be
h(\phi)=\phi-\phi^3+\frac35 \phi^5 - \frac 17 \phi^7.
\ee 
The first order equations reduce to $\phi^\prime=\pm(1-\phi^2)$, which can be solved by $\phi(x)=\pm \tanh(x)$. The energy density is $\rho(x)={\rm sech}^8(x)$ and the total energy is $E=32/35$. This is an example where the solution has the standard kink-like profile. See, e. g., Refs.~\cite{b1,b2} for more details on kinks in the presence of generalized dynamics.  

Another example is given by  
\be
h(\phi)=\frac{\phi(1-\phi^2)^{\frac32}}{4}+3\frac{\phi(1-\phi^2)^{\frac12}}{8}+\frac38 \arcsin(\phi).
\ee
Here the first-order equations reduce to $\phi^\prime=\pm (1-\phi^2)^{1/2}$. The solutions are  
\be
\phi(x)=\pm \biggl{\{} 1, \;\;\; x>\frac\pi2;\;\; \sin(x), \;\;\; -\frac\pi2<x<\frac\pi2; \;\;-1, \;\;\; x<-\frac\pi2\biggr{\}}.
\ee
The energy density is 
\be
\rho(x)=\pm \biggl{\{} 0, \;\;\; x>\frac\pi2;\;\; \cos^4(x), \;\;\; -\frac\pi2<x<\frac\pi2;\;\;0, \;\;\; x<-\frac\pi2\biggr{\}},
\ee
and the total energy is $E=3\pi/8$. This example is different, and all the solutions are compactons. This kind of solutions are growing in importance, in particular, in the study of pattern formation, since patterns usually appear in nature with finite extent, see, e.g., Refs.~\cite{s1,s2,b2,s3,c} for more details on compactons. {It is interesting to see that compactons
also appear in the supersymmetric context, so they are not incompatible with supersymmetry.}

The next step is to derive the fermion dependent part of the complete action composed by the sum of (\ref{sact}) and (\ref{pot1}). We get
\bea
\label{ferm00}
S_1^f&=&-\frac{1}{2}\int d^3x\Big[\psi^{\alpha}\psi_{\alpha}[f_{XX}(X)\pa^n\psi^{\beta}\pa^m\psi_{\beta}\pa_m\phi\pa_n\phi+
\frac{1}{2}f_X(X)\pa^m\psi^{\beta}\pa_m\psi_{\beta}+\nonumber\\&+&f_X(X)\pa^mF\pa_m\phi-h_{\phi\phi}(\phi)]+\nonumber\\&+&
2(i\pa_{\beta\alpha}\phi-C_{\beta\alpha}F)\psi^{\alpha}\pa^m\psi^{\beta}\pa_m\phi f_X(X)+\frac{i}{2}\psi^{\alpha}\pa_{\alpha\beta}\psi^{\beta}f(X)
\Big].
\eea
Eliminating the auxiliary field from the action being the sum of (\ref{acf1}) and (\ref{ferm00}) we find
\bea
F=-\frac{h_{\phi}}{f(X)}-\frac{1}{2f(X)}\pa^m(\psi^{\alpha}\psi_{\alpha}f_X(X)\pa_m\phi)-\frac{f_X(X)}{f(X)}\psi^{\alpha}\pa^m\psi_{\alpha}\pa_m\phi.
\eea
We can substitute this expression to the complete action and derive the equation of motion for the spinor field. This equation is highly nonlinear, but, if we choose the ansatz $\psi_{\alpha}=C_{\alpha}\chi(x)$ for the spinor field, with $C_{\alpha}$ as a Grassmannian constant, the nonlinear terms in $\psi_{\alpha}$ (or, as is the same, in $C_{\alpha}$) would vanish. Therefore we can restrict ourselves to the linearized equation which looks like
\bea
\label{lineq}
&&[h_{\phi\phi}(\phi)+f_X\pa^m(\frac{h_{\phi}(\phi)}{f(X)})\pa_m\phi]\psi_{\alpha}- i\pa_{\beta\alpha}\phi \pa^m\psi^{\beta}\pa_m\phi f_X(X)-\frac12 i\pa^m(\pa_{\alpha\beta}\phi\,\psi^{\beta}\pa_m\phi f_X(X))- \nonumber\\
&-&
\frac{h_{\phi}(\phi)}{f(X)}\pa^m\psi_{\alpha}\pa_m\phi f_X(X)+\frac12\pa^m\left(\frac{h_{\phi}(\phi)}{f(X)}\psi_{\alpha}\pa_m\phi f_X(X)\right)-\nonumber\\&-&\frac{i}{2}[\pa_{\alpha\beta}\psi^{\beta}f(X)+\pa_{\alpha\beta}(f(X)\psi^{\beta})]=0.
\eea

For static solutions, the zero mode can be found. In this case, we rewrite the Eq.~(\ref{lineq}) as   
\bea
\label{lineq1}
&&\left[h_{\phi\phi}(\phi)+f_X\left(\frac{h_{\phi}(\phi)}{f(X)}\right)^\prime\phi^\prime\right]\psi_{\alpha}+
\left[-\frac{h_{\phi}(\phi)}{f(X)}\psi^\prime_{\alpha}\phi^\prime f_X(X)+\frac12\left(\frac{h_{\phi}(\phi)}{f(X)}\psi_{\alpha}\phi^\prime f_X(X)\right)^\prime\right]-\nonumber\\&-&\frac{i}2 \gamma_{\beta\alpha}^1 \left[2\phi^{\prime2}\psi^{\beta\prime} f_X + (\phi^{\prime2} \psi^\beta f_X)^\prime+2\psi^{\beta\prime}f(X)+2\psi^{\beta}(f_X)\phi^\prime\phi^{\prime\prime}\right]=0.
\eea
We can apply the equation (\ref{firstorder}) which yields 
\bea
\label{lineq1a}
&&\left[h_{\phi\phi}(\phi)+f_X\phi^{\prime\prime}\phi^\prime\right]\psi_{\alpha}+
\left[-\psi^\prime_{\alpha}\phi^{\prime2} f_X(X)+\frac12\left(\psi_{\alpha}\phi^{\prime2} f_X(X)\right)^\prime\right]-\nonumber\\&-&\frac{i}2 \gamma_{\beta\alpha}^1 \left[2\phi^{\prime2}\psi^\beta f_X + (\phi^{\prime2} \psi^\beta f_X)^\prime+2\psi^{\beta\prime}f(X)+2\psi^{\beta\prime}(f_X)\phi^\prime\phi^{\prime\prime}\right]=0.
\eea
We choose $\gamma^1=\sigma_1$ and 
\be
\psi_\beta=\left(\begin{array}{c}
A(x) \\ 0
\end{array}\right)\, , \;\;\;\;
\psi^\alpha=\left(\begin{array}{c}
0 \\ -iA(x)\end{array}\right) \, , 
\ee
thus
\bea
\label{lineq1c}
&&h_{\phi\phi}(\phi)A=
A^\prime\left(
 2\phi^{\prime2} f_X +f(X)\right).
\eea
The solution of this equation is $A=\phi^\prime$. This result is not a surprise since in a supersymmetric theory the fermionic zero mode would exactly match the bosonic zero mode, and we know that the bosonic zero mode is the derivative of the static solution, a fact that follows from translational invariance of the theory.

We can also search for some other solutions. Unfortunately, even in the linearized case this equation is very complicated, especially due to the fact that while some terms after choice of the ansatz $\psi_{\alpha}=C_{\alpha}\chi(x)$ are proportional to $C_{\alpha}$, other ones are proportional to $C^{\beta}\gamma^m_{\beta\alpha}$. In particular, in the case $f(X)=1$ the equation (\ref{lineq}) changes to 
\bea
\label{lineq2}
&&h_{\phi\phi}(\phi)\psi_{\alpha}-%\nonumber\\&+&
i\pa_{\alpha\beta}\psi^{\beta}=0.
\eea
This is the Dirac equation which can be exactly solved for a diversity of cases.

Let us now introduce another theory, described by the following superfield action
\bea
S_2=-\frac{1}{4}\int d^5z D^{\alpha}\Phi D_{\alpha}\Phi f(\pa^m\Phi\pa_m\Phi) g(\Phi).
\eea
First, we will obtain its component structure. Similarly to the previous case, the bosonic sector of this theory can be found to have the form
\bea
S_2=-\frac{1}{4}\int d^3x D^2[D^{\alpha}\Phi D_{\alpha}\Phi] f(\pa^m\Phi\pa_m\Phi)g(\Phi)]|,
\eea
since terms where derivatives acting on $f(\pa^m\Phi\pa_m\Phi)$ or $g(\Phi)$ do not contribute to the bosonic sector because they evidently depend on fermions. Thus, we arrive at the following bosonic action
\bea
\label{bos}
S_2=-\frac{1}{2}\int d^3x (\pa^m\phi\pa_m\phi-FF) f(\pa^n\phi\pa_n\phi)g(\phi).
\eea
The equations of motion for the auxiliary field $F$ again yield $F=0$; thus, we arrive at the following action
\bea
S_2^b=-\frac{1}{2}\int d^3x \pa^m\phi\pa_m\phi f(\pa^n\phi\pa_n\phi)g(\phi),
\eea
or
\bea
S_2^b=-\frac{1}{2}\int d^3x Xf(X)g(\phi),
\eea
which for a special choice of $f(X)$ can yield Born-Infeld-like action.

The fermion dependent part of the action $S_2$ can be found to be equal to
\bea
\label{ferm}
S_2^f&=&-\frac{1}{2}\int d^3x\Big[\psi^{\alpha}\psi_{\alpha}[f_{XX}(X)g(\phi)\pa^n\psi^{\beta}\pa^m\psi_{\beta}\pa_m\phi\pa_n\phi+
\frac{1}{2}f_X(X)g(\phi)\pa^m\psi^{\beta}\pa_m\psi_{\beta}-\nonumber\\&-&\frac{1}{2}f(X)g_{\phi}(\phi)F+f_X(X)g(\phi)\pa^mF\pa_m\phi]+\nonumber\\&+&
2(i\pa_{\beta\alpha}\phi-C_{\beta\alpha}F)\psi^{\alpha}\pa^m\psi^{\beta}\pa_m\phi f_X(X)g(\phi)+\frac{i}{2}\psi^{\alpha}\pa_{\alpha\beta}\psi^{\beta}f(X)g(\phi)
\Big].
\eea
We eliminate the auxiliary field $F$ via its equations of motion obtained for the complete action  $S_2^{com}=S_2^b+S_2^f$ being the sum of (\ref{ferm}) and (\ref{bos}):
\bea
\label{f}
F&=&-\frac{1}{2f(X)g(\phi)}[2\psi^{\alpha}\pa^m\psi_{\alpha}\pa_m\phi f_X(X)g(\phi)+\frac{1}{2}\psi^{\alpha}\psi_{\alpha}f(X)g_\phi(\phi)+\nonumber\\&+&\pa^n(\psi^{\alpha}\psi_{\alpha}f_X(X)g(\phi)\pa_n\phi)].
\eea
It remains to substitute (\ref{f}) to the complete action. We get it in the form
\bea
S_2^{com}&=&-\frac{1}{2}\int d^3x \Big[ Xf(X)g(\phi)+\\&+&\psi^{\alpha}\psi_{\alpha}[f_{XX}(X)g(\phi)\pa^m\psi^{\beta}\pa^n\psi_{\beta}\pa_m\phi\pa_n\phi+
\frac{1}{2}f_X(X)g(\phi)\pa^m\psi^{\beta}\pa_m\psi_{\beta}-\nonumber\\&-&
\frac{4(f_X(X))^2g(\phi)}{f(X)}\pa^m\psi^{\beta}\pa^n\psi_{\beta}\pa_m\phi\pa_n\phi]+
\nonumber\\&+&2i\pa_{\beta\alpha}\phi\psi^{\alpha}\pa^m\psi^{\beta}\pa_m\phi f_X(X)g(\phi)+\frac{i}{2}\psi^{\alpha}\pa_{\alpha\beta}\psi^{\beta}f(X)g(\phi)
\Big].\nonumber
\eea
First, we can derive the equations of motion for the scalar field in the case of zero spinors. We get
\bea
-2\pa_n[(f(X)+Xf'(X))g(\phi)\pa^n\phi]+Xf(X)g_\phi(\phi)=0.
\eea
It is clear that the case $\phi=const$ (and hence $X=0$) is a possible solution of this equation. Another static solution is $\phi(x)=k_nx^n$, which corresponds to $X=k^2$. Defining ${\tilde f}(X)= X\,f(X)$ we get
\bea
{\tilde f}(X) - 2X   {\tilde f}_X(X) = 0.
\eea

We can also take
\be
\tilde f(X) = (1 + 2X)^a,
\ee
where $a$ is a positive parameter. In this case, we get to 
\be
(1+2X)^{a-1} \left[ 1+ 2(1-2a)X \right]=0.
\ee
Now, for $a>1/2$, we can write
\be
\phi=\frac{x}{\sqrt{2(2a-1)}}.
\ee
More details about this kind of solutions can be found in Ref.~\cite{b1}.

In summary, we have used the superfield formalism to obtain supersymmetric models engendering generalized dynamics. In particular, we have shown how to get to models which naturally lead to BPS states with their standard features, and how to find kinks and compactons in the presence of generalized dynamics. Even though we are considering much more complicated models, supersymmetry, once again, guide us toward the construction of generalized models where the calculations can be done analytically, in a simplified environment.

{The present results may be used to further explore the nature of the spacetime in the presence of compact and non-compact backgrounds, following the lines of \cite{mc,o,b2,CGB}. Further investigations are under consideration, in particular, on the search for other generalized models in flat spacetime, and on issues of direct interest to cosmology and braneworld. Specific investigations on how compact structures evolve cosmologically will be reported elsewhere.}

The authors would like to thank CAPES and CNPq for partial financial support.


\begin{thebibliography}{100}
\bb{V}A. Vilenkin and E. P. S. Shellard, {\it Cosmic strings and other topological defects} (Cambridge UP, Cambridge, UK, 1994).
\bb{BPS}E. B. Bogomol'nyi, Sov. J. Nucl. Phys. {\bf24}, 449 (1976); M. K. Prasad and C. M. Sommerfield, Phys. Rev. Lett. {\bf35}, 760 (1975).
\bb{m1}C. Armendariz-Picon, T. Damour, and V. Mukhanov, Phys. Lett. B {\bf458}, 209 (1999) [hep-th/9904075].
\bb{m2}C. Armendariz-Picon, V. Mukhanov, and P. J. Steinhardt, Phys. Rev. Lett. {\bf85}, 4438 (2000) [astro-ph/0004134];
C. Armendariz-Picon, V. Mukhanov, and Paul J. Steinhardt, Phys. Rev. D {\bf63}, 103510 (2001) [astro-ph/0006373].
\bb{mc}M. Cvetic, S, Griffies, and S.-J. Rey, Nucl. Phys B {\bf381}, 301 (1992) [hep-th/9201007]; M. Cvetic and H. H. Soleng, Phys. Rep. {\bf282}, 159 (1997) [hep-th/9604090].
\bb{B}E. Babichev, Phys. Rev. D {\bf74}, 085004 (2006) [hep-th/0608071].
\bb{b1}D. Bazeia, L. Losano, R. Menezes, and J. C. R. Oliveira, Eur. Phys. J. C {\bf51}, 953 (2007) [hep-th/0702052].
\bb{s1}C. Adam, J. Sanchez-Guillen, and A. Wereszczynski, J. Phys. A {\bf40}, 13625 (2007) [arXiv:0705.3554]; Erratum-ibid. A {\bf42}, 089801 (2009).
\bb{o}M. Olechowski, Phys. Rev. D {\bf78}, 084036 (2008) [arXiv:0801.1605].
\bb{s2}C. Adam, N. Grandi, J. Sanchez-Guillen, and A. Wereszczynski, J. Phys. A {\bf41}, 212004 (2008) [arXiv:0711.3550];
C. Adam, N. Grandi, P. Klimas, J. Sanchez-Guillen, and A. Wereszczynski, J. Phys. A {\bf41}, 375401 (2008) [arXiv:0805.3278];
C. Adam, P. Klimas, J. Sanchez-Guillen, and A. Wereszczynski, J. Phys. A {\bf42}, 135401 (2009) [arXiv:0811.4503].
\bb{b2}D. Bazeia, L. Losano, and R. Menezes, Phys. Lett. B {\bf668}, 246 (2008), [arXiv:0807.0213]; D. Bazeia, A. R. Gomes, L. Losano, and R. Menezes, Phys. Lett. B {\bf671}, 402 (2009) [arXiv:0808.1815].
\bb{s3}C. Adam, P. Klimas, J. Sanchez-Guillen, and A. Wereszczynski, {\it Compact shell solitons in K field theories}, [arXiv:0902.0880]; C. Adam, N. Grandi, P. Klimas, J. Sanchez-Guillen,
and A. Wereszczynski, {\it Compact boson stars in K field theories} [arXiv:0908.0218].
\bb{c}P. Rosenau and J. M. Hyman, Phys. Rev. Lett. {\bf70}, 564 (1993); P. Rosenau, Phys. Rev. Lett. {\bf73}, 1737 (1994); P. Rosenau and A. Pikovski, Phys. Rev. Lett. {\bf94}, 174102 (2005); P. Rosenau, J. M Hyman, and M. Staley, Phys. Rev. Lett. {\bf98}, 024101 (2007).
\bibitem{CGB}D. Z. Freedman, C. Nunez, M. Schnabl, and K. Skenderis, Phys. Rev. D {\bf69}, 104027 (2004) [hep-th/0312055];
D. Bazeia, C. B. Gomes, L. Losano, and R. Menezes, Phys. Lett. B {\bf633}, 415 (2006) [astro-ph/0512197]; V. I. Afonso, D. Bazeia, and L. Losano, Phys. Lett. B {\bf634}, 520 (2006) [hep-th/0601034]; K. Skenderis and P. K. Townsend, Phys. Rev. Lett. {\bf96}, 191301 (2006) [hep-th/0602260];  D. Bazeia, L. Losano, J.J. Rodrigues, and R. Rosenfield, Eur. Phys. J. C {\bf55}, 113 (2008) [astro-ph/0611770]; D. Bazeia, B. Carneiro da Cunha, R. Menezes, and A. Yu. Petrov, Phys. Lett. B {\bf649}, 445 (2007) [hep-th/0701106]; E. A. Bergshoeff, J. Hartong, A. Ploegh, J. Rosseel, and D. Van den Bleeken, JHEP {\bf 0707}, 067 (2007) [arXiv:0704.3559]; K. Skenderis, P. K. Townsend, and A. Van Proeyen, JHEP {\bf0708}, 036 (2007) [arXiv:0704.3918]; D. Bazeia, F. A. Brito, and F. G. Costa, Phys. Lett. B {\bf661}, 179 (2008) [arXiv:0707.0680]; D. Bazeia, A. R. Gomes, and L. Losano, Int. J. Mod. Phys. A {\bf24}, 1135 (2009) [arXiv:0708.3530];
V. I. Afonso, D. Bazeia, R. Menezes, and A. Yu. Petrov, Phys. Lett. B {\bf658}, 71 (2008) [arXiv:0710.3790];  M. Cvetic and M. Robnik, Phys. Rev. D {\bf77}, 124003 (2008) [arXiv:0801.0801];
D. Bazeia, R. Menezes, and A. Yu. Petrov, Eur. Phys. J. C {\bf58}, 171 (2008) [arXiv:0806.2299]; A. de Souza Dutra, A.C. Amaro de Faria Jr., and M. Hott, Phys. Rev. D {\bf78} 043526 (2008) [arXiv:0807.0586]; C. A. S. Almeida, M. M. Ferreira Jr., A. R. Gomes, and R. Casana, Phys. Rev. D {\bf79} 125022 (2009) [arXiv:0901.3543]; P.~P.~Avelino, D.~Bazeia, L.~Losano, R.~Menezes, and J.~J.~Rodrigues, Phys.\ Rev.\  D {\bf 79}, 123503 (2009) [arXiv:0903.5297]; Yu-Xiao Liu, Jie Yang, Zhen-Hua Zhao, Chun-E Fu, and Yi-Shi Duan, Phys. Rev. D {\bf80} 065019 (2009) [arXiv:0904.1785]; Yu-Xiao Liu, Chun-E Fu, Li Zhao, and Yi-Shi Duan, Phys. Rev. D {\bf80} 065020 (2009) [arXiv:0907.0910].
\bibitem{SGRS} S. J. Gates, M. T. Grisaru, M. Rocek, and W. Siegel. {\it Superspace or One Thousand and One Lessons in Supersymmetry} (Benjamin/Cummings, NY, 1983).
\bibitem{my3}H. O. Girotti, M. Gomes, A. Yu. Petrov, V. O. Rivelles, and A. J. da Silva. Phys. Lett. B {\bf521}, 119 (2001) [hep-th/0109222]; H. O. Girotti, M. Gomes, A. Yu. Petrov, V. O. Rivelles, and A. J. da Silva, Phys. Rev. D {\bf67}, 125003 (2003) [hep-th/0207220].
\bibitem{ours}I. L. Buchbinder, S. M. Kuzenko, and J. V. Yarevskaya, Nucl. Phys. B {\bf411}, 665 (1994); I. L. Buchbinder, S. M. Kuzenko, and A. Yu. Petrov, Phys. Lett. B {\bf321}, 372 (1994); Phys. At. Nucl. {\bf59}, 148 (1996).
\end{thebibliography}
\end{document}